\documentclass[aps,pre,showpacs,preprintnumbers,amsmath,amssymb,11 pt]{revtex4}
\usepackage[dvips]{graphicx}
\usepackage{amsmath}
\usepackage{latexsym}
\usepackage{amsthm}
\usepackage{amssymb}
\usepackage{color}
\usepackage[sort&compress]{natbib}

\begin{document}
\title{Crowding effects in non-equilibrium transport through nano-channels}

\author{A. Zilman$^{1,2}$, G. Bel$^{2,3}$}
\affiliation{
$^{1}$Theoretical Biology and Biophysics Group and $^{2}$Center for Nonlinear Studies,
Theoretical Division\\
$^{3}$Computer, Computational and Statistical Sciences Division\\
Los Alamos National Laboratory, Los Alamos, NM 87545, USA }
\begin{abstract}
Transport through nano-channels plays an important role in many biological processes and industrial applications.
Gaining insights into the functioning of biological transport processes and the design of man-made nano-devices requires an understanding of the basic physics of such transport.
A simple exclusion process has proven to be very useful in explaining the properties of several artificial and biological nano-channels. It is particularly useful for modeling the influence of inter-particle interactions on transport characteristics. In this paper, we explore several models of the exclusion process using a mean field approach and computer simulations. We examine the effects of crowding inside the channel and its immediate vicinity on the mean flux and the transport times of single molecules.
Finally, we discuss the robustness of the theory's predictions with respect to the crucial characteristics of the hindered diffusion in nano-channels that need to be included in the model.
\end{abstract}
\maketitle

\section{Introduction}
Molecular transport at the nano-scale lies at the core of many biological processes and novel technological applications \cite{non-ehxaustive,alberts-book,nanofluidics-review-2005,stein-book-1990}.
Biological examples of such transport include numerous molecular transporters, such as porins, ion channels, various secretion systems, nucleo-cytoplasmic exchange, and many other cellular channels
and transporters \cite{schulten_glycerol,bezrukov-zwitterion-OMPF,click-mitochnodrial-import-1998,rout-review-2003,tran-wente-review-2006,stewart-review-2007,aquaporins-book,bezrukov-kullman-maltoporin-2000,eisenberg-OMPF-2004}.
These biological devices transport molecules into and out of the cell, as well as between different cellular compartments.
This transport is highly regulated  and selective. Such transporters function as 'molecular filters' that transport only their cognate molecules while efficiently filtering out everything else.

Nano-channels also play an increasingly important role in technological and engineering applications.
Examples include nano-fluidics, novel biosensors, nano-molecular sorters, nano-pore based DNA sequencing and molecular separation processes \cite{kasianowicz-2008-review,kasianowicz-2001-protein-pore,martin-DNA-2004,martin-apoenzymes-1997,nanofluidics-review-2005,polymer-nanotubules-2008,akin-DNA-nanotubes-2007,ziwy-gillespie-anomanlus-mole-fraction-BJ-2008,caspi-elbaum-2008,non-ehxaustive}. The artificial channels can serve as model systems to mimic and explain the principles of similar biological channels \cite{caspi-elbaum-2008,jovanovic-nature-2009}.

Modern experimental techniques provide a wealth of information about the function of such devices and an opportunity to directly test the theoretical concepts.
Several experimental methods are common in the study of the transport through nano-channels.
On the macroscopic scale, measurements of the static and dynamic behavior of fluxes through the channel are well established with a variety of methods \cite{stein-book-1990,rout-review-2003,timney-mike-JCB-2006,akin-DNA-nanotubes-2007,bezrukov-kullman-maltoporin-2000,ziwy-gillespie-anomanlus-mole-fraction-BJ-2008,bezrukov-antibiotics-PNAS-2002,bezrukov-BJ-mito,non-ehxaustive}.
On a microscopic scale, single molecule tracking techniques can measure directly the transport times and their distributions
\cite{yang-musser-JCB-2006,musser-single-PNAS-2004, kubitscheck-JCB-2008,kubitscheck-JCB-2005}.
The various techniques provide information about different aspects of the transport process.

The mechanisms of the function of such channels are still not fully understood. In order to fully realize the potential of nano-channel based technologies and in order to understand the biological transport properties, one needs to understand the basic physics
of transport through nano-channels--from macroscopic fluxes down to single molecules.
Despite their apparent biological and technological complexity and diversity, it appears that the function of many such transporters can be understood from basic principles within
simplified models that use a small set of coarse-grained parameters and capture the essentials of the problem. These models are commonly based on non-equilibrium statistical mechanics \cite{redner-book,gardiner-book-2003,van-kampen-book,non-ehxaustive,kolomeisky-2006,chou-PRL-single-file-1998,chou-zeolites-PRL-1999,zilman-BJ-2009,zilman-plos-2007,bezrukov-ptr-2002,bezrukov-optimal-2005,bezrukov-asymmetric-2007,berezhkovskii-hummer-PRL-2002,zilman-pearson-bel-PRL-2009,bauer-nadler-pnas-2006,cussler-book-1997,ziwy-gillespie-anomanlus-mole-fraction-BJ-2008,eisenberg-diffusion-discrete-1988,eisenberg-barriers-2007}.
Analytical results obtained from simplified models provide important insights into the basic principles of the system and enable one to investigate separately the effects of different factors in idealized situations, un-confounded by other factors. In addition, such models guide the creation of new artificial channels and can direct one how to manipulate the channel properties in order to achieve a desired functionality. When necessary, the simple models can be generalized to include more details, such as charge and avidity effects \cite{eisenberg-barriers-2007,gillespie-eisenberg-2002-dft,gillespie-eisenberg-dft-2003,chou-tasep-2003}, and they may be complemented and compared with molecular dynamics, Brownian dynamics and other simulation techniques \cite{berezhkovskii-hummer-PRL-2002,simulatinos-review-2008,ziwy-gillespie-anomanlus-mole-fraction-BJ-2008,gillespie-eisenberg-Monte-Carlo}.

In general, the choice of a model hinges on the balance between computational feasibility and conceptual transparency on one hand, and the ability to capture all the essential richness of the experimental setup on the other. One important feature that has to be captured in a model of transport through nano-channels is the non-equilibrium crowding and interactions between the molecules inside the channel. In particular, the exclusion process model of the confinement and coarse-grained interactions of the particles with the channel has proved to be a good starting point \cite{schutz-review-2005,chou-PRL-single-file-1998,macdonald-gibbs-1968,derrida-mukamel-exclusion-1992,zilman-BJ-2009,zilman-plos-2010}. It has been successfully applied to the explanation of various processes, such as the transport of DNA through nano-channels and the nuclear pore complex \cite{zilman-BJ-2009,zilman-plos-2010}.

In this paper, we expand our previous models \cite{zilman-pearson-bel-PRL-2009,zilman-BJ-2009} of the hindered diffusion in nano-channels as an exclusion process to include the channel asymmetry and non-uniformity, in order to capture more realistically the diffusion and binding processes near the channel entrance. The analytical results are validated against kinetic Monte Carlo simulations. Finally, we discuss what aspects of the model are sensitive to the inclusion of these features and the applicability of the model to the analysis of experimental data.

\section{The Model}
\begin{figure}[htbp]
\centerline{
\includegraphics[width= 8 cm]{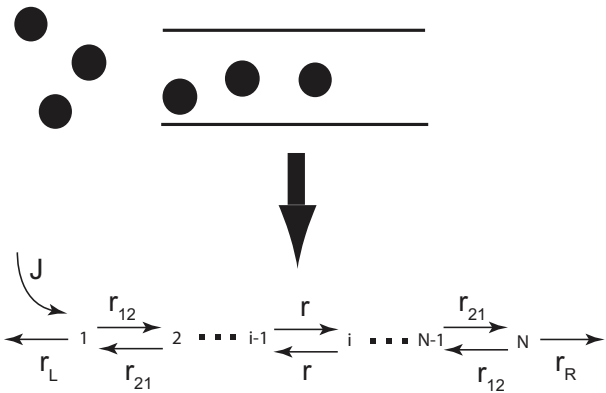}
}
\caption{\label{fig-schematic} Schematic illustration of the mapping of transport through a nano-channel onto an exclusion process.
The rates $r_L$ and $r_R$ represent the diffusion from the immediate vicinity of the channel entrance towards the bulk.
$r_{21}$ and $r_{12}$ represent the exit/entrance rates and the dynamics in the immediate vicinity of the channel.
The rate $r$ represents the diffusion inside the channel. The particles enter at site $1$ and hop between adjacent sites if their occupancy is less than the maximal occupancy $m$,
which models the steric hindrance.}
\end{figure}

In the exclusion process model, the channel is represented as a sequence of $N$ 'sites'. It is important to emphasize that these 'sites' do not represent the actual binding sites of the molecules inside the channel, but rather serve as a convenient computational tool to
model the hindered diffusion. The particles enter from the left with constant flux $J$ and can hop to the adjacent sites if their occupancy is less than the maximal occupancy $m$, which models the crowding and competition for space. The interactions of the particles with the channel are represented by the rates of hopping through and the exit from the channel. This dynamics is illustrated in Fig.~\ref{fig-schematic}. The model allows the calculation of both the macroscopic properties as the fluxes and transport probabilities through the channel as well as properties which characterize the single molecules being transported as the average translocation/return times and their distributions.

\section{Uniform asymmetric channel}
As a simplest approximation, we first discuss a uniform channel, where the dynamics at the immediate vicinity of the channel is not considered explicitly ($r_{12}=r_{21}=r$, see Fig.~\ref{fig-schematic}); we discuss the implications of the different choices of the rates in Section \ref{Sec-NU}.

\subsubsection{Single Particle}
We first discuss the transport of a single particle through the channel, in the absence of the impinging flux $J$.
The particle starts at the 'entrance' site $1$ and can hop to the right to site $2$, with rate $r$, or hop to the left, out of the channel, with rate $r_L$. From any internal site $1<i<N$, the particle can hop either left or right with equal probability, with rate $r$ (for each direction). From the last site $N$, the particle can hop either to the right, out of the channel, with rate $r_R$ or to the left, to site $N-1$, with the rate $r$. The strength of the interaction of the particle with the channel is modeled by the choice of $r_{L}$ and $r_R$.
It is important to note that particles can hop out of the channel in both directions and do not always exit on the right.

The motion of the  particle through the channel is described by the time dependent vector of probabilities $p_i(t)$ to be at a particular site $i$: $|p(t)\rangle=(p_1(t),...p_i(t)...p_N(t))$, so that $\langle i|p(t)\rangle= p_i(t)$.

The Master equation for the probability vector $|p(t)\rangle$, describing the hopping through and out of the channel ends, is \cite{redner-book,zilman-pearson-bel-PRL-2009,bezrukov-asymmetric-2007,van-kampen-book}
\begin{equation}\label{equation-for-p-1-prl}
\frac{d}{dt}|p(t)\rangle=\hat{M}\cdot |p(t)\rangle,
\end{equation}
where the matrix $\hat{M}$ has the following elements:
\begin{eqnarray}
M_{i,i}&=&-2r,\;\;\;M_{i,i\pm 1}=r\;\;\text{for}\;\;1<i<N\\
M_{1,1}&=&-r-r_L,\;\;\;M_{N,N}=-r-r_R,\;\;\;M_{1,2}=M_{N,N-1}=r.\nonumber
\end{eqnarray}
It is useful to introduce the Laplace transform of the probabilities vector $|\tilde{p}(s)\rangle={\displaystyle\int\limits_0^\infty}e^{-s t}|p(t)\rangle dt$.
In Laplace space, the Master equation \eqref{equation-for-p-1-prl} becomes
\begin{equation}
 s|\tilde{p}(s)\rangle-|p(t=0)\rangle=\hat{M}\cdot|\tilde{p}(s)\rangle.
\end{equation}
The solution to this equation is $|\tilde{p}(s)\rangle=(s\hat{I}-\hat{M})^{-1}\cdot|p(t=0)\rangle $. Or, in explicit form:
\begin{equation}\label{sol1}
 \tilde{p}_{i}(s)=A(\lambda_1^i-\left(\frac{\lambda_1}{\lambda_2}\right)^{N-1}\frac{\left(r+s+r_R\right)\lambda_1-r}{\left(r+s+r_R\right)\lambda_2-r}\lambda_2^i),
\end{equation}
where
\begin{equation}\label{sol2}
 \lambda_{1,2}=\frac{2r+s\pm\sqrt{s^2+4rs}}{2r},
\end{equation}
and
\begin{equation}\label{sol3}
 1/A=-\left(-\left(r+r_L+s\right)\lambda_1+r\lambda_1^2+\left(\frac{\lambda_1}{\lambda_2}\right)^{N-1}\frac{-r+\left(r+r_R+s\right)\lambda_1}{-r+\left(r+r_R+s\right)\lambda_2}\left(r+r_R+s-r\lambda_2\right)\right).
\end{equation}
The expressions above for the Laplace transform enable one to calculate various quantities of interest, such as the transport probabilities and the moments of the transport times.

The instantaneous probability flux to the right out of the channel is $r_Rp_N(t)$. Thus, the probability that the particle exited the channel from the right side by time $t$ is simply the accumulated probability flux $P_{\rightarrow}^t=\int_0^t r_Rp_N(t')dt'$. This is what is often measured in bulk flux or single molecule experiments \cite{yang-musser-JCB-2006}.

The total probability to exit to the right is
\begin{equation}\label{equation-Pr-single-prl}
P_{\rightarrow}\equiv P_{\rightarrow}^{\infty}=\int_0^{\infty}r_R\langle N|e^{\hat{M}t}|1\rangle dt'=r_R \tilde{p}_{N}(s=0).
\end{equation}
Using the expression for $\tilde{p}_{i}(s)$, we get the translocation probability as
\begin{equation}
 P_{\rightarrow}=\frac{r r_R}{\left(N-1\right)r_L r_R+r\left(r_L+r_R\right)}.
\end{equation}
As expected, for $r_L=r_R$ (i.e., for a symmetric channel), this result reduces to known results \cite{zilman-pearson-bel-PRL-2009,bezrukov-asymmetric-2007}.
In the limit $r_R\gg r_L$ and fixed $r$, $P_{\rightarrow}\rightarrow 1$ and when $r_L\gg r_R$ and $r$ is fixed, $P_{\rightarrow}=r_R/r_L\to 0$, as expected.
In the limit of fast diffusion inside the channel, when both $r_L,r_R\ll r$, one gets $P_{\rightarrow}=r_R/\left(r_R+r_L\right)$; that is, the channel essentially behaves as a one-site channel \cite{bezrukov-sites-2005}.
Note that for an asymmetric channel the translocation probability varies between $0$ and $1$, unlike the case of a symmetric channel where the upper limit for the translocation probability is $1/2$.

We now turn to the calculation of the transport times. The probability density of the exit times to the right $f_{\rightarrow}(t)$ is the derivative of the cumulative distribution $P_{\rightarrow}^t$, divided by the total probability to exit to the right.
In Laplace space, this is expressed as $f_{\rightarrow}(s)=\frac{r_R \tilde{p}_{N}(s)}{r_R \tilde{p}_{N}(s=0)}$.
Therefore, the mean first passage time to exit to the right is
\begin{equation}
\overline{T}_{\rightarrow}=-\frac{1}{\tilde{p}_{N}(s=0)}\frac{d\tilde{p}_{N}(s)}{ds}\Big|_{s=0}.
\end{equation}
Similarly, the mean first passage time to exit to the left is
\begin{equation}
\overline{T}_{\leftarrow}=-\frac{1}{\tilde{p}_{1}(s=0)}\frac{d\tilde{p}_{1}(s)}{ds}\Big|_{s=0}.
\end{equation}
The mean exit time from \textit{any} of the ends is
\begin{eqnarray}
\overline{T}=-\left(r_L\frac{d\tilde{p}_{1}(s)}{ds}+r_R\frac{d\tilde{p}_{1}(s)}{ds}\right)\Big|_{s=0}.
\end{eqnarray}
Using the explicit solution for the probabilities, from Eqs. (\ref{sol1}),(\ref{sol2}),(\ref{sol3}), we get:
\begin{eqnarray}
\overline{T}_{\rightarrow}&=&\frac{N \left(\left(N^2-3 N+2\right)r_L r_R+3 (N-1) r (r_L+r_R)+6 r^2\right)}{6 r ((N-1)r_L r_R+r(r_L+r_R))}\nonumber\\
\overline{T}_{\leftarrow}&=&\frac{N \left(\left(2 N^2-3 N+1\right) r_R^2+6 (N-1) r r_R+6 r^2\right)}{6 ((N-1) r_R+r) ((N-1) r_L r_R+r(r_L+r_R))}\nonumber\\
\overline{T}&=&\frac{N}{2r}\frac{2+\left(N-1\right)r_R/r}{\left(N-1\right)r_L r_R/r^2+\left(r_L+r_R\right)/r}.
\end{eqnarray}
Similar results were obtained in \cite{berezhkovskii-times-2003}.

\begin{figure}[htbp]
\centerline{
\includegraphics[width=0.8\linewidth]{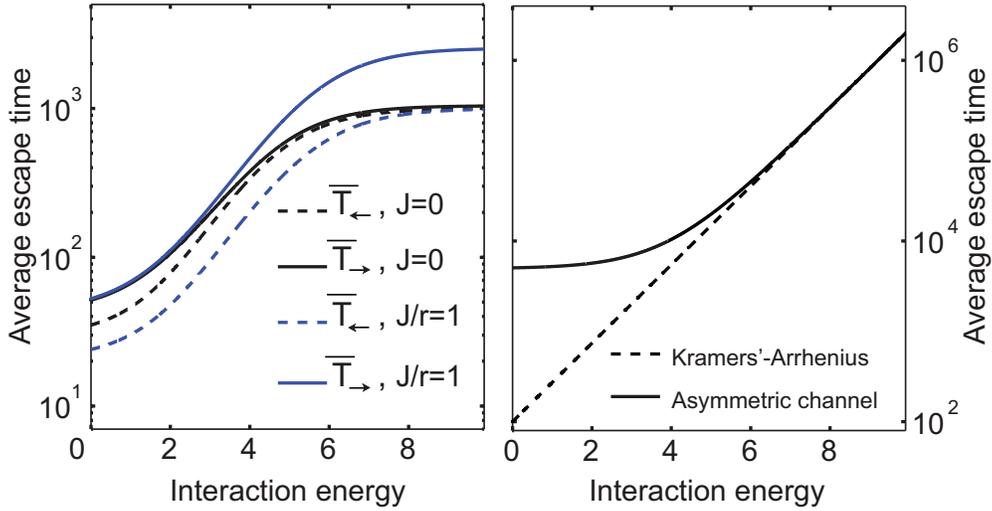}
}
\caption{\label{fig-mean-times-kramers} \emph{Left panel}: Average escape times from a channel versus the interaction energy $E=-\ln(r_R)$ for $r_L=0.01$. The solid lines correspond to the average forward escape time and the dashed lines to the average backward escape times. The black lines correspond to $J=0$, namely the average escape times of a single particle and the blue lines correspond to $J/r=1$ where many particles are crowded in the channel. The  channel length is $N=10$. \emph{Right panel}: Comparison with the Kramers-Arrhenius theory. The solid line is the uni-directional escape time from a potential well as a function of the interaction energy $E=-\ln(r_R)$ for $r_L=0$ and $J=0$. The dashed line is the exponential asymptote of the Arrhenius-Boltzmann-Kramers type.}
\end{figure}
The behavior of the mean transport times is illustrated in Fig.~\ref{fig-mean-times-kramers}. The right panel shows the escape time from a potential well with a reflective boundary condition on the left ($r_L=0$). In the  regime of high interaction energy $E=-\ln(r_R)$, the dependence of the escape time asymptotically reaches the exponential $\exp(-E)$ in accord with the heuristic expectation of the Arrhenius-Boltzmann-Kramers' type \cite{gardiner-book-2003,van-kampen-book}.
By contrast, when there is a finite escape rate to the left ($r_L\neq 0$), the left panel in Fig.~\ref{fig-mean-times-kramers} shows that the directional average escape time to the right does not follow the exponential asymptote as a function of the binding energy $E$, but rather saturates to a constant, determined by the value of the exit rate to the left, $r_L$.

\subsection{Non-equilibrium flux through the channel: crowding}
\subsubsection{Steady state flux}
When a finite flux $J$ impinges onto the channel entrance, a non-equilibrium steady state is established, so that at any time there can be many particles in the channel that might interfere with each other's passage. The particles in the channel obey the same kinetics as described in the previous section, with the condition that a site can contain up to a maximal number of particles $m$. One can describe the system in terms of the average site occupancies $n_i$. After the relaxation of the initial condition, a steady-state density profile $|n\rangle^{ss}$ is established.
For the case of a uniform channel
\begin{equation}\label{equation-ni}
n^{ss}_i=\frac{J (r_R (N-i)+r)}{J ((N-1)r_R+r)/m+(N-1)r_Lr_R+r (r_L+r_R)}.
\end{equation}
The average exit flux to the right is $J_{\rightarrow}=r_Rn^{ss}_N$, which yields for the probability of an individual particle within this flux to exit to the right
\begin{equation}\label{p-exit-ss-prl}
P_{\rightarrow}=\frac{J_{\rightarrow}}{J(1-n^{ss}_1/m)}=\frac{r r_R}{\left(N-1\right)r_L r_R+r\left(r_L+r_R\right)}.
\end{equation}
Note that the translocation probability of individual particles is the same as in the single-particle case (at least for uniform channels), even though the particles are interfering with each other's passage through the channel.
Another characteristic of the transport is the efficiency which measures the fraction of the impinging flux which is being transported through the channel:
\begin{equation}
\text{Eff}\equiv\frac{J_{\rightarrow}}{J}=\frac{r_R r}{J ((N-1)r_R+r)/m+(N-1)r_Lr_R+r (r_L+r_R)}.
\end{equation}
Similar results have been obtained in \cite{chou-PRL-single-file-1998,rittenberg-1996,derrida-mukamel-exclusion-1992,zilman-pearson-bel-PRL-2009}.

Unlike in a symmetric channel, the exit rates at both ends can vary independently, as they can be determined by different physical mechanisms. Moreover, the crowding can also influence the diffusion rate $r$ inside the channel (for instance, it has been observed in several systems that the \emph{on} and \emph{off} rates of molecular binding can change due to crowding \cite{martin-DNA-2004}). These and other effects lead to a rich behavior of the steady state flux through the channel.

\begin{figure}[htbp]
\centerline{
\includegraphics[width=0.8\linewidth]{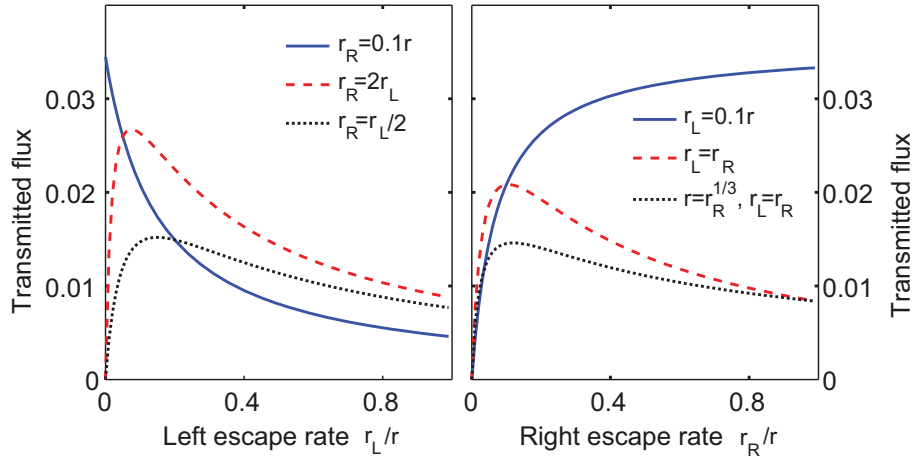}
}
\caption{\label{fig-flux-vs-rate} Transmitted flux through a channel of length $N=10$ versus the escape rates $r_L$, $r_R$. The impinging flux is $J/r=0.1$ for all lines except the black dashed line in the right panel for which $J=0.1$ in the arbitrary units used for the other rates (in this line the hopping rate inside the channel $r$ depends on the hopping rate outside the channel $r_R$).}
\end{figure}

\begin{figure}[htbp]
\centerline{
\includegraphics[width=0.8\linewidth]{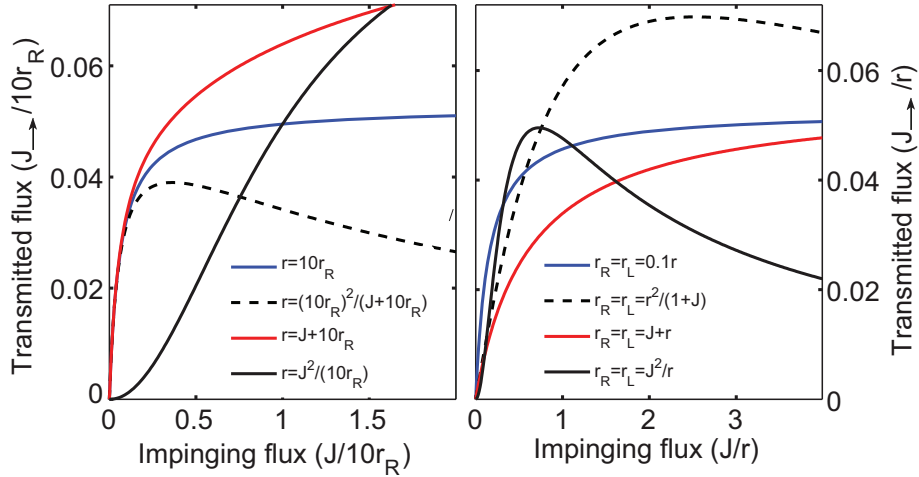}
}
\caption{\label{fig-flux-vs-impinging-flux}
Transmitted flux through a channel of length $N=10$ versus the impinging flux.
The kinetic rates can depend on the impinging flux; see text.
\emph{Left panel}: $r_R=0.1$, $r_L=0.01$.
\emph{Right panel}: $r=1$. }
\end{figure}

The general qualitative features of the flux behavior are illustrated in Figs.~\ref{fig-flux-vs-rate},\ref{fig-flux-vs-impinging-flux}. In particular, they show that in the case when the exit rates vary independently from each other, the transmitted flux is a monotonic function of the rates (Fig.~\ref{fig-flux-vs-rate}). Only in the case when the values of the exit rates are connected (of which a symmetric channel is a special case) via a common physical mechanism, the flux (or the efficiency) exhibits optimum with  respect to the exit rates (Fig.~\ref{fig-flux-vs-rate}). Thus, varying the relative strength of trapping at the channel ends, it is possible to tune the selectivity properties of the channel, which has important implications for the design of artificial nano-channels.

The dependence of the transmitted flux on the impinging flux $J$ is illustrated in Fig.~\ref{fig-flux-vs-impinging-flux}. The left panel of Fig.~\ref{fig-flux-vs-impinging-flux} shows the dependence of the transmitted flux on the impinging flux when the diffusion rate inside the channel depends on the impinging flux (the different lines correspond to different dependence of the internal diffusion on the impinging flux) and the rates of hopping from the channel outside are fixed. When the diffusion rate inside the channel is independent of the impinging flux or is a monotonically increasing function of it, the transmitted flux exhibits the usual saturation curve. Such a situation can arise when the effective binding  energies  decrease due to inter-particle crowding \cite{martin-DNA-2004,stewart-nup-binding-to-imps-2002}. On the other hand, when the diffusion rate inside the channel is a monotonically decreasing function of the impinging flux there is an optimal value for the impinging flux which maximizes the transmitted flux. Such a situation can arise experimentally when the transported molecules change the conformation and structure of the channel constituents \cite{yang-musser-JCB-2006}.

The crowding can also have an effect on the exit rates. In the right panel of Fig.~\ref{fig-flux-vs-impinging-flux}, we show the dependence of the transmitted flux on the impinging flux when the exit rates out of the channel directly depend on the impinging flux. Similarly to the case where the flux influences the diffusion inside, the transmitted flux exhibits an optimum as a function of the impinging flux both for the case in which the exit rates are a monotonically decreasing function of the impinging flux and for the case in which they depend quadratically on the impinging flux. When the exit rates are independent of or increase with the impinging flux the optimum disappears, and the transmitted flux monotonically increases and saturates as a function of the impinging flux.

\subsubsection{Single particle on the background of the steady state flux}\label{sec:Sec-Crowding-times}
In order to interpret single-molecule tracking experiments in the experimentally relevant regimes,  we investigate how the transport times of the individual particles are influenced by the crowding, when many particles are present in the channel. To the best of our knowledge, no closed analytical solution exists in this case. We use the mean field approximation, backed by kinetic Monte Carlo simulations.

To this end, transport of an individual particle can be viewed as occurring on the background of the steady state density profile $|n\rangle^{ss}$. In the mean field (or the effective medium) approximation, the probability $p_i(t)$ of a particle to be present at a given site is described by the following equations \cite{zilman-pearson-bel-PRL-2009}:
\begin{eqnarray}\label{kinetics_with_exclusion-ss-prl}
\frac{d}{dt}p_i&=&rp_{i-1}(1-n^{ss}_i/m)+rp_{i+1}(1-n^{ss}_i/m)-rp_i(1-n^{ss}_{i-1}/m)-rp_i(1-n^{ss}_{i+1}/m),
\end{eqnarray}
with the boundary conditions
\begin{align}\label{kinetics_with_exclusion-ss-prl-bc}
\frac{d}{dt}p_1&=rp_{2}(1-n^{ss}_1/m)-rp_1(1-n^{ss}_{2}/m)-r_Lp_1,\\\nonumber
\frac{d}{dt}p_N&=rp_{N-1}(1-n^{ss}_N/m)-rp_N(1-n^{ss}_{N-1}/m)-r_Rp_N.
\end{align}
Similar to the case of a single particle, this set of equations can be written in a matrix form:
\begin{equation}\label{kinetics_with_exclusion-ss-prl-Matrix}
\frac{d}{dt}|p(t)\rangle=\hat{M}_{ss}\cdot|p(t)\rangle,
\end{equation}
with the matrix elements of $\hat{M}_{ss}$ following from Eqs.(\ref{kinetics_with_exclusion-ss-prl},\ref{kinetics_with_exclusion-ss-prl-bc}).
In Laplace space, this becomes
\begin{equation}
|\tilde{p}(s)\rangle=(s\hat{I}-\hat{M}_{ss})^{-1}\cdot|p(t=0)\rangle.
\end{equation}
\subsubsection{Mean transport times}
\begin{figure}[htbp]
\centerline{
\includegraphics[width=0.8\linewidth]{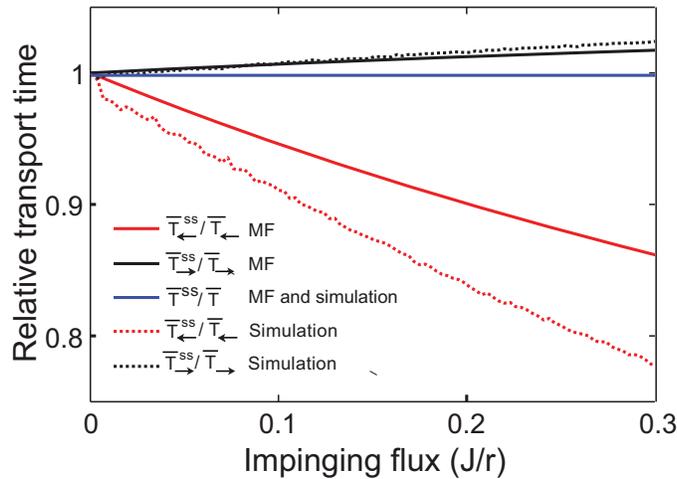}
}
\caption{\label{fig-mean-times-vs-sim} Dependence of the mean exit times on the impinging flux $J$, normalized to the case with $J=0$.
Blue middle line: the average dwelling time $\overline{T}^{ss}_{\rightarrow}/\overline{T}_{\rightarrow}$.
 Lower line: the return time $\overline{T}^{ss}_{\leftarrow}/\overline{T}_{\leftarrow}$.
  Upper line: the forward exit time $\overline{T}^{ss}_{\rightarrow}/\overline{T}_{\rightarrow}$.
   The corresponding simulation results are plotted with dotted lines.
   The parameters are: $N=6$, $r=1$, $m=3$, $r=1$, $r_L=0.01$, $r_R=0.1$. }
\end{figure}

Equations (\ref{kinetics_with_exclusion-ss-prl}),(\ref{kinetics_with_exclusion-ss-prl-bc}) and (\ref{kinetics_with_exclusion-ss-prl-Matrix}) can be solved in a manner similar to the single particle case above. In particular, the translocation probability turns out to be the same as for a single particle in the absence of flux, as has already been established in the previous sections (based on the ratio between the transmitted and entering fluxes). After some algebra, one gets for the mean time to exit to the left:
\begin{widetext}
\begin{eqnarray}
\overline{T}^{ss}_{\leftarrow}=
-\frac{
N\left[\frac{2 r m}{J P_{\rightarrow}}+4\frac{r}{r_R}+3\left(N-1\right)\right]
-2 \left(\frac{r m}{J P_{\rightarrow}}+\frac{r}{r_R}+N-1\right)^2 \left[\psi\left(N+\frac{m r}{J P_{\rightarrow}}\right)-\psi\left(\frac{m r}{J P_{\rightarrow}}\right)\right]}
{2 J \left(N-1+\frac{r}{r_R}\right)/m}.
\end{eqnarray}
\end{widetext}
where $\psi(x)=\frac{d}{dx}\Gamma(x)$; $\Gamma(x)$ is a $\gamma$-function. A similar expression can be obtained for $T^{ss}_{\rightarrow}$ (not shown).
The mean time to exit to the right can be calculated from the relation
\begin{equation}
P_{\rightarrow}\overline{T}^{ss}_{\rightarrow}+P_{\leftarrow}\overline{T}^{ss}_{\leftarrow}=\overline{T}^{ss}.
\end{equation}
Interestingly, the mean escape also turns out to be unaffected by the crowding, similar to the uniform channel case \cite{zilman-pearson-bel-PRL-2009} (see Fig.~\ref{fig-mean-times-vs-sim}).

The dependence of the mean escape times on the impinging flux is shown in Fig.~\ref{fig-mean-times-vs-sim}. Similarly to the symmetric channel case, the crowding increases the forward exit time and decreases the backward exit time. The mean field approximation is compared with the kinetic Monte Carlo simulations (dotted lines in Fig. \ref{fig-mean-times-vs-sim}), which shows that it is an excellent approximation for wide channels ($m=3$) and is a reasonable approximation also in the case of single file transport ($m=1$) even for quite high values of the impinging flux $J$, where the crowding effects are significant.

Fig.~\ref{fig-mean-times-kramers} (right panel) illustrates the dependence of the average escape times to the left and to the right on the interaction energy. Interestingly, the crowding does not change the qualitative behavior of the escape times.

It is also interesting to note that
\begin{equation}
J(1-n^{ss}_1)\overline{T}^{ss}=\mathbb{N}_{occup},
\end{equation}
where $\mathbb{N}_{occup}=\sum_{i=1}^{i=N}n_i$ is the average number of particles in the channel. Heuristically, despite the crowding effects, one can think of the channel as a 'box' with $\mathbb{N}_{occup}$ particles that hop out with an average rate $1/\overline{T}^{ss}$ so that the total flux out of the channel is $J_{\rightarrow}+J_{\leftarrow}=\mathbb{N}_{occup}/\overline{T}^{ss}$. This heuristics is very different from the one applied to the potential well with a uni-directional escape (see Kramers' theory). For instance, it is not true that $J_{\rightarrow}=\mathbb{N}_{occup}/\overline{T}^{ss}_{\rightarrow}$. It is important to note that the above mentioned heuristics is only valid when thinking about the average flux; higher moments will deviate from that picture since the distribution of exit times is not exponential.

\subsubsection{Probability distributions of the transport times}
\begin{figure}[htbp]
\centerline{
\includegraphics[width=0.8\linewidth]{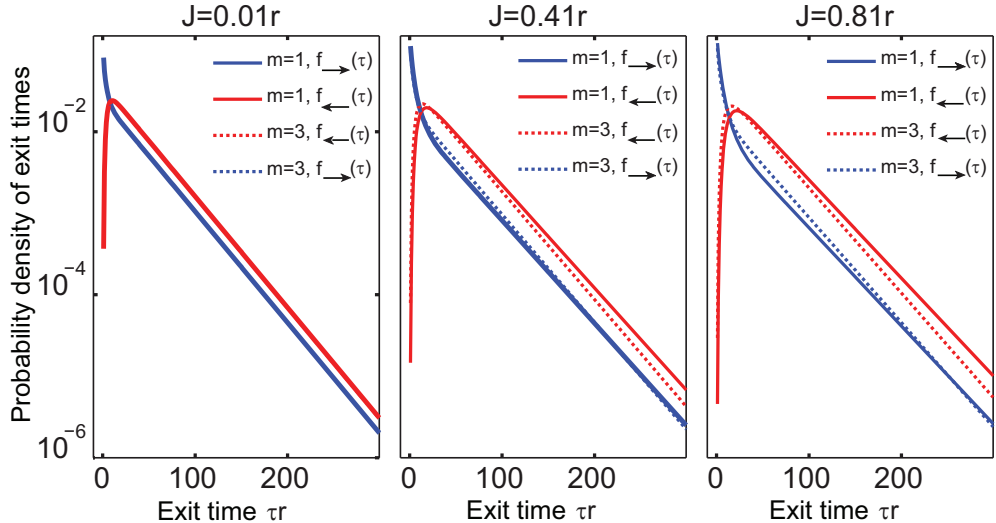}
}
\caption{\label{fig-times-dist-MF}
The distributions of both the forward and the backward  exit times decay exponentially
with the same constant at long times. The parameters are shown in each panel. In all panels $N=6$, $r=1$, and $r_L=r_R=0.1$.  }
\end{figure}

\begin{figure}[htbp]
\centerline{
\includegraphics[width=0.9\linewidth]{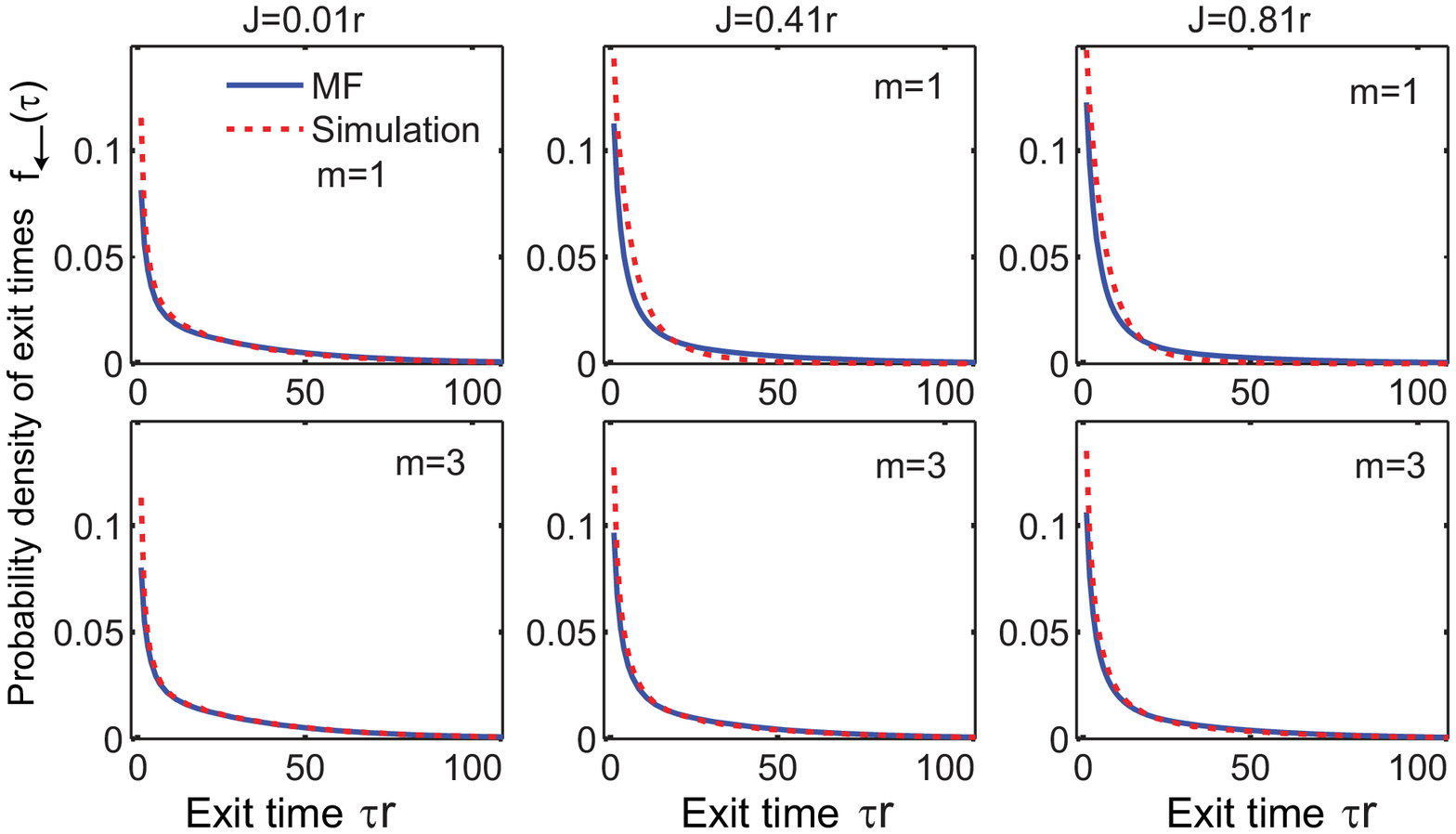}
}
\caption{\label{fig-bcktimes-dist} Backward exit time distributions for different values of $J$.
 Solid blue line: mean field results.
 Dotted red line: simulations.
 \emph{Upper panels}: m=1;\emph{Lower panels}: m=3. In all panels $N=6$, $r=1$, and $r_L=r_R=0.1$. }
\end{figure}

\begin{figure}[htbp]
\centerline{
\includegraphics[width=0.8\linewidth]{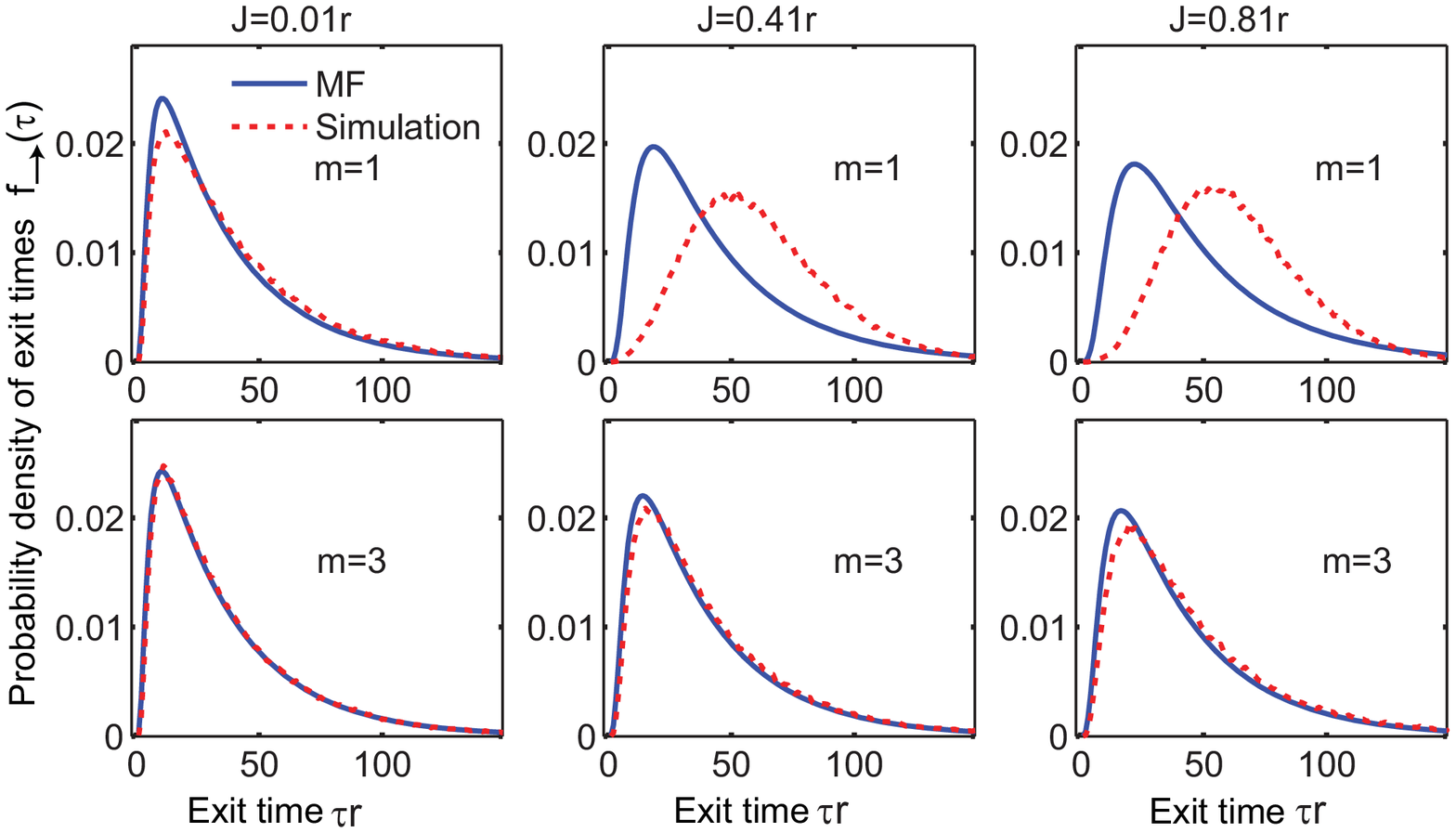}
}
\caption{\label{fig-fwtimes-dist}Forward exit time distributions for different values of $J$.
 \emph{Upper panels}: m=1; \emph{Lower panels}:  m=3.
 In all panels $N=6$, $r=1$, and $r_L=r_R=0.1$. }
\end{figure}

The single molecule tracking experiments can measure not only the mean transport times, but also the transport time distributions \cite{yang-musser-JCB-2006,kubitscheck-JCB-2005,kubitscheck-JCB-2008}, which can be calculated within the present formalism. As mentioned above, the distribution of the exit times  to the right $f_\rightarrow(t)$ is simply $r_R p_N(t)/P_{\rightarrow}$, where $p_N$ is the probability that the tagged particle is in site $N$ at time $t$ and the distribution of the exit times to the left is $f_\leftarrow(t)=r_L p_1(t)/P_{\leftarrow}$ \cite{redner-book,gardiner-book-2003,zilman-pearson-bel-PRL-2009,berezhkovskii-times-2003}. Thus, from equations~(\ref{kinetics_with_exclusion-ss-prl}, \ref{kinetics_with_exclusion-ss-prl-bc}, \ref{kinetics_with_exclusion-ss-prl-Matrix}), the probability distribution of the exit times to the right of a particle that enters the channel at site $1$ is
\begin{equation}
f_\rightarrow(t)=\frac{r_R}{P_{\rightarrow}}\exp\left(\hat{M}_{ss} t\right)_{N1}.
\end{equation}
In Laplace space, we can write it as $\tilde{f}_\rightarrow(s)=(r_R/P_{\rightarrow})\langle N|(s\hat{I}-\hat{M}_{ss})^{-1}|p(t=0)\rangle$. In the single particle limit, $J\rightarrow 0$, when the $n_{ss}\rightarrow 0$ and $\hat{M}_{ss}\rightarrow \hat{M}$, it reduces to the expressions of the previous section. The probability distributions of the transport times are illustrated in Figs.~\ref{fig-times-dist-MF}, \ref{fig-fwtimes-dist} and \ref{fig-bcktimes-dist}.

Both the forward and the backward exit times have an exponential decay tail (as demonstrated by the straight lines in Fig.~\ref{fig-times-dist-MF}). Note that the decay constant is the same for both forward and backward times (as demonstrated by the fact that the lines for the forward and backward exit times probability density are parallel in Fig.~\ref{fig-times-dist-MF}), because the tails of the distributions are due to particles that have been bouncing back and forth inside the channel for a long time and thus have lost the memory of the initial condition.

The distribution of the return time is strongly non-mono-exponential. The decay at very short times is much faster than at long times. This can be understood by noting that the decay at very short times is due to those returning particles that jump out immediately after entering the channel. Therefore, the short time decay rate of the probability density of the left exit times is $r+r_L$. It is important to note that even in the limit of simple diffusion, when $r_L=r_R=r$, the distribution of the backward exit times is still non-mono-exponential; this bears important implications for the interpretation of single molecule tracking experiments \cite{kubitscheck-JCB-2008,yang-musser-JCB-2006}.

Comparison of the mean field probability density with the simulations in Figs.~\ref{fig-bcktimes-dist} and~\ref{fig-fwtimes-dist} shows that for $m=3$ the mean field and simulation curves are almost identical for all values of $J$. For the single file case, $m=1$, and large values of $J$ there is a small difference between the mean field and the simulation curves due to the fact that the mean field fails to capture correlations between successive hops for strictly single file transport in the crowded regime \cite{zilman-pearson-bel-PRL-2009}.



\section{Non-uniform channels: particles returning to the channel \label{Sec-NU}}
The uniform channel is one of the simplest models that captures the essentials of hindered diffusion with interactions through a narrow channel. It has been shown to account well for the effects of the inter-particle interactions on the flux \cite{zilman-BJ-2009}, and some of its more complicated predictions in the multi-species case have been verified experimentally \cite{zilman-plos-2010,jovanovic-nature-2009}. An interesting and useful feature of the uniform channel model is that many quantities of interest are independent of the impinging flux and are un-affected by the crowding \cite{zilman-BJ-2009,schutz-review-2005,zilman-pearson-bel-PRL-2009}.

However, when thinking of a uniform channel as a model of diffusion through a channel, it is apparent that it neglects an important effect: the possibility that a particle can return into the channel after it has exited. Moreover, the diffusion rate in the immediate vicinity of the channel entrance can differ from both the diffusion rate in the bulk and inside the channel. We now model the dynamics near the channel entrance in more detail, by accounting for the possibility that the  particles can return into the channel after they exited it. This effect is modeled by choosing the sites $1$ and $N$ to lie 'outside' the channel. This is illustrated in Fig.~\ref{fig-schematic}. Now the rate $r_{21}$ represents the escape rate from the channel to the medium in the immediate vicinity of the channel end, while the rate $r_{12}$ represents the return rate into the channel. Accordingly, $r_L$ and $r_R$ represent free diffusion from the immediate vicinity of the channel towards the bulk (to the left and right correspondingly). As before, $r$ is the diffusion rate inside the channel and $J$ is the flux impinging from the bulk. The sites outside the channel ($1,N$) are assumed to be free of any restriction on the number of particles they can contain.

The local exit and entry rates $r_{21}$ and $r_{12}$ contain the potentially complicated dynamics of binding/unbinding and local diffusion near the entrance. From the detailed balance condition, their ratio is equal to the Boltzmann factor of the energy difference $E$ between the inside and the outside of the channel: $r_{21}/r_{12}=\exp(-E)$. However, their actual values are determined by the local diffusion coefficient, which in turn depends on the details of the energy profile of the molecular interactions near the channel entrance.

The mean field equations for the steady state densities in this case are \cite{macdonald-gibbs-1968,van-kampen-book,schutz-review-2005,chou-tasep-2003}
\begin{equation}\label{eq-steady-state-n-nonuniform}
0=-2rn_{i}^{ss}+r\left(n_{i+1}^{ss}+n_{i-1}^{ss}\right) \ \ \ \ \text{for}\ \ \ \ 3\leq i \leq N-2,
\end{equation}
and for the boundary sites:
\begin{align}
J&=r_Ln^{ss}_{1}+r_{12}n^{ss}_{1}\left(1-n^{ss}_{2}/m\right)-r_{21}n^{ss}_{2} \nonumber \\
0&=-r_{21}n_{2}^{ss}-rn_{2}^{ss}(1-n_{3}^{ss}/m)+r_{12}n_{1}^{ss}(1-n_{2}^{ss}/m)+rn_{3}^{ss}(1-n_{2}^{ss}/m) \nonumber \\
0&=-r_{21}n_{N-1}^{ss}-rn_{N-1}^{ss}(1-n_{N-2}^{ss}/m)+r_{12}n_{N}^{ss}(1-n_{N-1}^{ss}/m)+rn_{N-2}^{ss}(1-n_{N-1}^{ss}/m) \nonumber \\
0&=-r_{R}n_{N}^{ss}-r_{12}n_{N}^{ss}(1-n_{N-1}^{ss}/m)+r_{21}n_{N-1}^{ss}.
\end{align}
These can be easily solved by noticing that the mean field equations for the densities at  the internal sites are identical to the equations describing the steady state population in uniform channels: $n_{i}^{ss}=a+bi$ for $2\leq i \leq N-1$. The set of equations~(\ref{eq-steady-state-n-nonuniform}) can then  be solved for the variables $a, b, n^{ss}_{1}$ and $n^{ss}_{N}$. The explicit solution has been obtained in Mathematica but since it is cumbersome it will be omitted here.

\begin{figure}[htbp]
\centerline{
\includegraphics[width=0.8\linewidth]{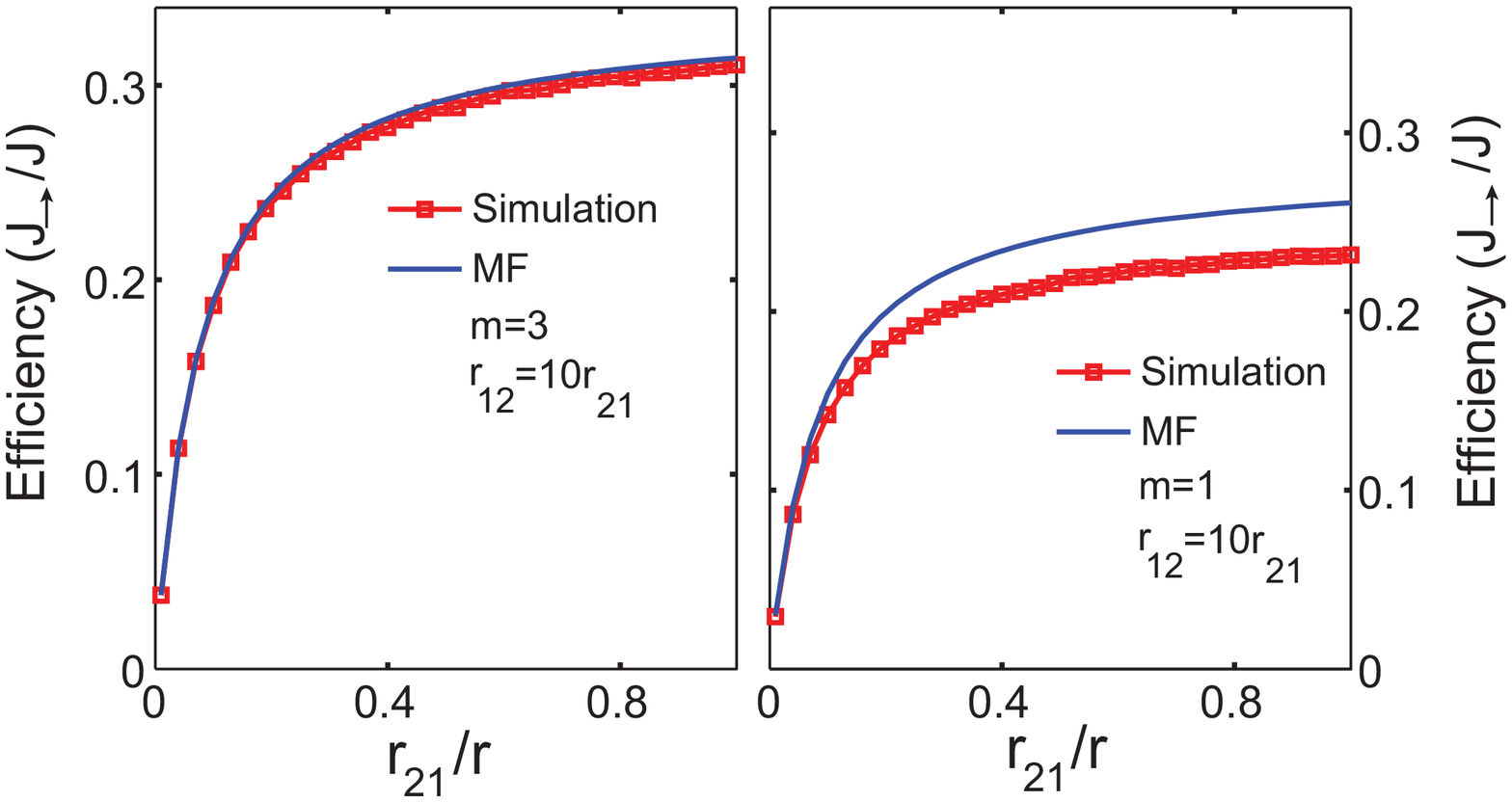}
}
\caption{\label{fig-Eff-D}The transport efficiency $J_{\rightarrow}/J$ versus the local diffusion
coefficient at the channel ends for a fixed ratio $r_{21}/r_{12}=0.1$.
The blue lines show the analytical mean field results while the red lines show the simulations of
the exact dynamics. The left panel shows the case where the maximal occupancy of each site is $m=3$ and the
right panel shows the case of single file channel $m=1$. In all panels $J=0.1$, $r=r_L=r_R=1$, and $N=10$.}
\end{figure}
\begin{figure}[htbp]
\centerline{
\includegraphics[width=0.8\linewidth]{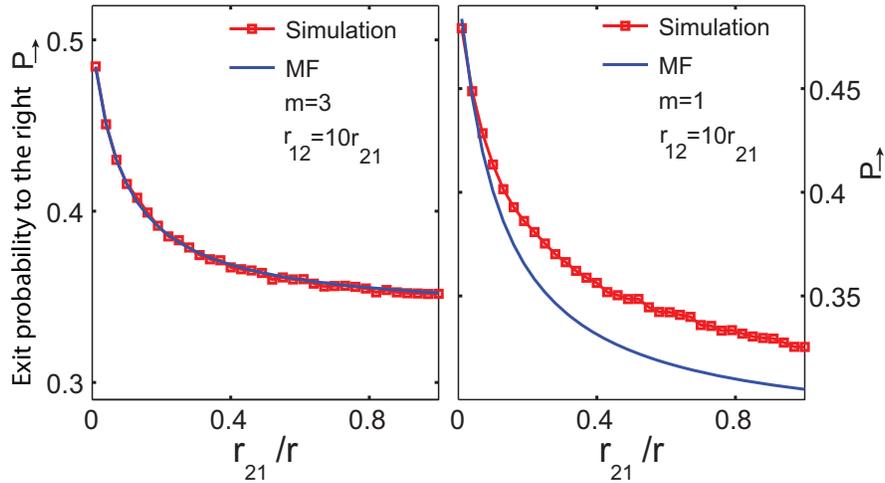}
}
\caption{\label{fig-PR-D} The translocation probability $P_{\rightarrow}$ versus the local diffusion
coefficient at the channel ends for a fixed ratio $r_{21}/r_{12}=0.1$.
The blue lines show the analytical mean field results while the red lines show the simulations of
the exact dynamics. The left panel shows the case where the maximal occupancy of each site is $m=3$ and the
right panel shows the case of single file channel $m=1$. In all panels $J=0.1$, $r=r_L=r_R=1$, and $N=10$.}
\end{figure}

The flux to the right through the channel is given by $J_{\rightarrow}=r_R n^{ss}_{N}$ and the transport efficiency is given by $J_{\rightarrow}/J$, which can be readily calculated from Eqs.~(\ref{eq-steady-state-n-nonuniform}). Not all the particles in the impinging flux are able to enter the channel, because the channel entrance  (site $2$) can be occupied. Thus, the transport probability of a particle to traverse the channel to the right once it has entered (into the site $2$) is $J_{\rightarrow}/J_{\text{in}}$, where $J_{\text{in}}=J\frac{r_{12}(1-n_{2}^{ss}/m)}{r_L+r_{12}(1-n_{2}^{ss}/m)}$ is the flux that actually enters the channel.

Figures~\ref{fig-Eff-D} and~\ref{fig-PR-D} show the transport efficiency and the translocation probability as a function of the rate $r_{21}$ for \emph{fixed} ratio  $r_{21}/r_{12}$. That is, in those figures the local diffusion coefficient at the channel entrance/exit varies, while binding energy stays constant. For wide channels (maximal occupancy $m=3$), there is an excellent agreement between the mean field calculations and the results of simulations of the exact dynamics. For $m=1$  (single file dynamics) the mean field still describes the efficiency reasonably well. The transport probability decreases with $r_{21}$, because larger $r_{21}$ implies an increased chance of hopping out of the channel. Somewhat counter-intuitively, the transport efficiency increases with $r_{21}$. This is due to the fact that for fixed ratio $r_{21}/r_{12}$, the larger $r_{21}$ implies larger $r_{12}$, which acts to keep the particle near the channel entrance instead of diffusing away to the bulk. These results indicate a possibility of an additional 'knob' for the control of transport properties of nano-channels.

\begin{figure}[htbp]
\centerline{
\includegraphics[width=0.8\linewidth]{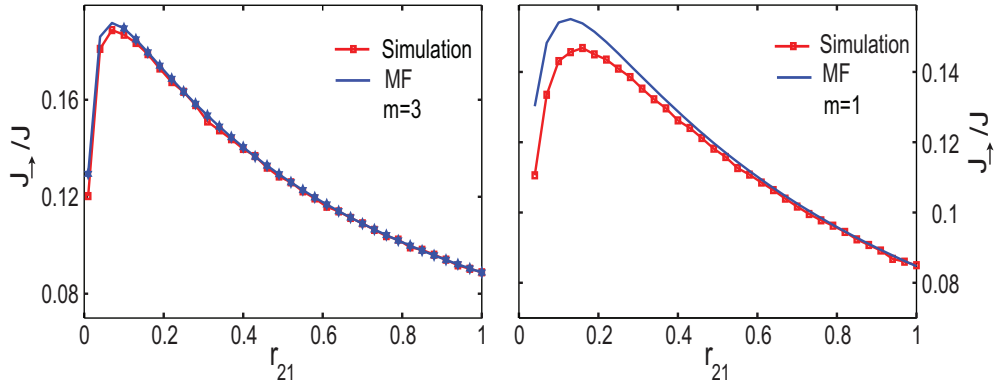}
}
\caption{\label{fig-Eff-dE}The transport efficiency $J_{\rightarrow}/J$ versus the binding energy at the channel ends. For these plots we have used the following parameters: $J=0.1$, $r=r_L=r_R=r_{12}=1$ and $N=10$. The left panel shows the case where the maximal occupancy of each site is $m=3$ and the right panel shows the case of single file channel $m=1$. The blue lines show the analytical mean field results while the red lines show the simulations of the exact dynamics.}
\end{figure}

\begin{figure}[htbp]
\centerline{
\includegraphics[width=0.8\linewidth]{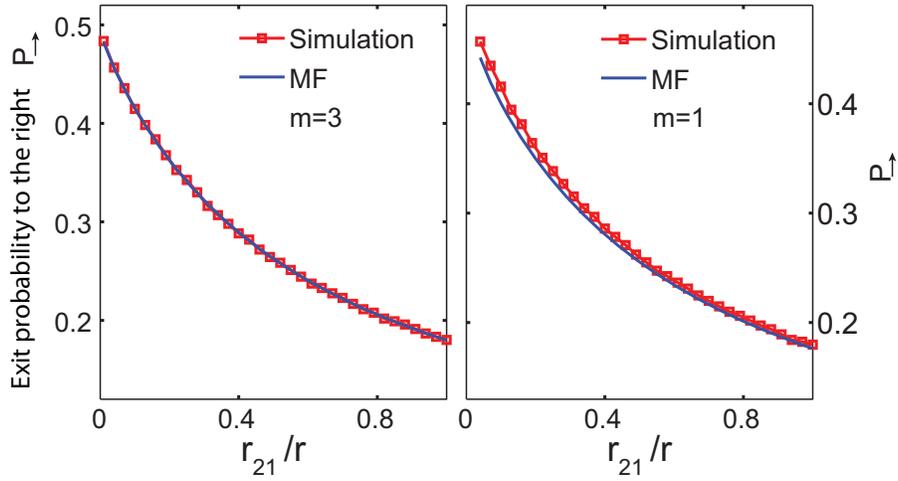}
}
\caption{\label{fig-PR-dE}The translocation probability versus the binding energy at the channel ends. The parameters are the same as in Fig.~\ref{fig-Eff-dE}. The left panel shows the case where the maximal occupancy of each site is $m=3$ and the right panel shows the case of single file channel $m=1$. The blue lines show the analytical mean field results while the red lines show the simulations of the exact dynamics.}
\end{figure}

Figures~\ref{fig-Eff-dE} and~\ref{fig-PR-dE} show the efficiency and the translocation probability versus $r_{21}$, but in the case where $r_{12}$ is fixed, that is as a function of the binding energy at the channel entrance. As above, the mean field approximation provides a good fit to the results of the simulations. However, unlike in Fig.~\ref{fig-Eff-D}, the transport efficiency is a non-monotonic function of $r_{21}$ and has an optimum at a certain value of $r_{21}$. In this respect, the non-uniform channel behaves as a uniform case that has been studied before \cite{chou-PRL-single-file-1998,zilman-BJ-2009}.


\subsection{Diffusion outside: choice of the model}
To what extent can the effect of the returning particles be neglected when analyzing the experimental data? Clearly, the non-uniform model, in which the sites outside the channel are explicitly considered, is more complicated and involves more parameters than the uniform channel model. It is important to know whether bulk measurements of the flux can differentiate between the two models, necessitating the explicit inclusion of the particle return for interpretation of the experimental data.

Here we start investigating this question by comparing the mean field exit flux through a symmetric non-uniform channel with length $N$ and rates $r_{12}, r_{21}, r_{R}=r_L, r$ with the exit flux through a uniform channel with the length $\tilde{N}$ and rates $r_o,r$. Qualitatively, both fluxes exhibit a saturating Michaelis-Menten type behavior as a function of the impinging flux $J$. It turns out that with the appropriate choice of the effective parameters $\tilde{N}$ and $r_o$ the behavior of the fluxes is also \emph{quantitatively} similar.
\begin{figure}[htbp]
\centerline{
\includegraphics[width=0.6\linewidth]{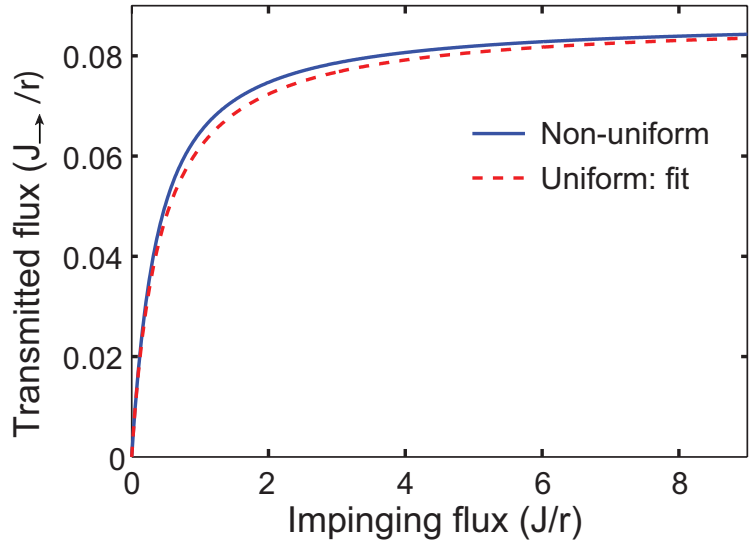}
}
\caption{\label{fig-Jout-fit}The output flux through a non-uniform (solid blue line) and an approximating uniform channel (dashed red line) versus the impinging flux $J$.
The parameters are  $r=r_R=r_L=1$, $r_{12}=2$, $r_{21}=0.5$ and $N=10$. The fitted parameters are $\tilde{N}=7.8$ and $r_o=0.3$, close to the naively expected values (see text for details).
}
\end{figure}

This result is illustrated in Fig.~\ref{fig-Jout-fit}, which  shows the exit fluxes versus the impinging flux for the nonuniform model and the corresponding uniform model. The effective length $\tilde{N}$ and the effective exit rate $r_o$ for the uniform channel were obtained by requiring that the exit flux at the saturation limit $J\to\infty$ and the slope of the exit flux with respect to $J$ at $J=0$ are equal for both models; these are two parameters that completely determine the shape of the Michaelis-Menten type curves. In Fig.~\ref{fig-Jout-fit}, the parameters for the non-uniform model are: $r=r_R=r_L=1$, $r_{12}=2$, $r_{21}=0.5$ and $N=10$. The fit of the uniform model yields the following parameters: $\tilde{N}=7.8$ and $r_o=0.3$, which are not too far from the effective parameters one would obtain by naively removing the sites outside the channel (in the equivalent uniform model) and setting the ratio $r_o/r=r_{21}/r_{12}$. The fitted curve is very close to the curve obtained from the non-uniform model (the deviation is much smaller than the experimental precision in the experiments that we are familiar with). The  agreement between the two models is rather general and does not depend on a fortuitous choice of parameters (data not shown), although its quality decreases when $r_{12},r_{21}\ll r$. Therefore, in general, in order to distinguish between the different models, additional experiments at the single molecule level are needed.

It is important to note that the two lines coincide at the limits $J\to 0$ and $J\to\infty$ (these are the fitting requirements) but still deviate in the intermediate regime. This deviation is a clear indication of a non Michaelis-Menten behavior of the output current for non-uniform channels. These effects lie outside of the scope of the present work and will be studied elsewhere. Finally, it is important to compare the models in the experimentally relevant regime. We have recently shown that the exclusion process model describes well the observed behavior of the flux of single-stranded DNA through nano-channels functionalized with DNA hairpins \cite{zilman-BJ-2009,martin-DNA-2004}. For  the parameter values arising from those experiments, the two models are essentially indistinguishable (data not shown) for the \emph{same} values of the physical parameters.

\section{Summary}
In this paper we have investigated the behavior of  the exclusion process models of non-equilibrium transport through nano-channels, using a mean field (effective medium) approach and kinetic Monte Carlo simulations. We have investigated the dependence of the static and dynamic characteristics of the transport on various parameters such as the binding energy in the channel, the diffusion rate inside and outside the channel and the flux through the channel in a uniform asymmetric channel. In particular, we have shown that the analytical mean field approximation captures well the properties of the transport both at the macroscopic level (fluxes, probabilities) and the single molecule level (transport times and their distributions). The results provide additional insights into how the properties of the artificial nano-channels can be manipulated to achieve a desired functionality.

We have also investigated an extended model which explicitly takes into account the dynamics of particles' exit and return into the channel in the immediate vicinity of its entrance.
For a wide range of parameters, in terms of describing the behavior of the steady state properties, such as fluxes and transport probabilities, inclusion of this effect does not
result in a difference that could be distinguishable on the basis of the measurement of the flux alone. However, the effects of local non-uniformity near the channel entrance might be observable in single-molecule experiments.

To summarize, simple physical models of the diffusive transport through nano-channels provide important insights into the mechanisms of their function. Even the simple exclusion models described in this paper have a rich behavior that appears to be capable of grasping the essential physics in several experimental systems. Nevertheless, the question of choice of the pertinent parameters that have to be included in the model for meaningful interpretation of the experiments still remains open and requires future works, both theoretical and  experimental, such as single molecule tracking measurements.

\acknowledgments{This work was performed under the auspices of the US Department of Energy under contract DE-AC52-06NA25396.}

\bibliographystyle{iop}
\bibliography{transport-May2010}

\begin{thebibliography}{10}
\bibitem{non-ehxaustive}
Due to scope and length limitations, the literature list is not exhaustive.
\end{thebibliography}

\end{document}